\documentclass{elsart3p}

\usepackage{graphicx}
\usepackage{epsfig}

\usepackage{amssymb}

\begin{document}

\begin{frontmatter}



\title{The fundamental role of superconducting quasiparticle
coherence in cuprate superconductors}


\author{Shiping Feng, Huaiming Guo}

\address{Department of Physics, Beijing Normal University, Beijing
100875, China}

\begin{abstract}
Within the kinetic energy driven superconducting mechanism, we
study the interplay between superconductivity and the nodal and
antinodal superconducting quasiparticle coherences in cuprate
superconductors, and find the s-wave superconducting transition
temperature is heavily suppressed by the antinodal superconducting
quasiparticle coherence, while the d-wave superconducting
transition temperature is enhanced, therefore the antinodal
superconducting quasiparticle coherence plays a more crucial role
in superconductivity of cuprate superconductors.
\end{abstract}

\end{frontmatter}

After twenty years intensive investigations, it has now become
clear that cuprate superconductors are among the most complex
systems studied in condensed matter physics \cite{kastner}. These
compounds have a layered structure of the square lattice of
CuO$_{2}$ plane separated by insulating layers, which leads to
that cuprate superconductors are highly anisotropic materials,
then the electronic excitation spectrum is dependent on the
in-plane momentum \cite{shen,fink}. The undoped cuprates are the
Mott insulators with an antiferromagnetic (AF) long-range order
(AFLRO), then upon the charge carrier doping, these compounds
evolve into the superconductors leaving the AF short-range
correlation still intact \cite{kastner}. Experimentally,
angle-resolved photoemission spectroscopy (ARPES) experiments have
made a great deal of progress in the understanding of the
underlying superconducting (SC) quasiparticle coherence of cuprate
superconductors. On the one hand, superconductivity in doped
cuprates results when electrons pair up into Cooper pairs
\cite{shen} as in the conventional superconductors. Although the
SC pairing mechanism is beyond the conventional electron-phonon
mechanism, the SC-state of cuprate superconductors \cite{matsui}
is the conventional Bardeen-Cooper-Schrieffer (BCS) like, so that
the basic BCS formalism is still valid in discussions of the
electron excitation spectrum. In particular, the information
revealed by ARPES experiments has shown that around the nodal and
antinodal points of the Brillouin zone contain the essentials of
the whole low energy quasiparticle excitation spectrum of cuprate
superconductors \cite{shen,fink}. On the other hand, unlike the
conventional superconductors, the SC quasiparticle coherence plays
an important role in superconductivity of doped cuprates
\cite{ding}. Although both nodal and antinodal SC quasiparticle
peaks exist for a wide range of the doping, the nodal and
antinodal SC quasiparticle coherent weights tend to zero at the
zero doping \cite{shen}. In this case, a natural question is
whether of the nodal \cite{johnson} and antinodal \cite{ding} SC
quasiparticle coherences plays a more crucial role for
superconductivity in doped cuprates.

Recently, we have developed a kinetic energy driven SC mechanism
\cite{feng1} based on the charge-spin separation (CSS)
fermion-spin theory \cite{feng2}, where the dressed holons
interact occurring directly through the kinetic energy by
exchanging the spin excitations, leading to a net attractive force
between the dressed holons, then the electron Cooper pairs
originating from the dressed holon pairing state are due to the
charge-spin recombination (CSR), and their condensation reveals
the SC ground-state. In particular, this SC-state is controlled by
both SC gap function and quasiparticle coherence, then the maximal
SC transition temperature occurs around the optimal doping, and
decreases in both underdoped and overdoped regimes \cite{feng3}.
In this paper, we discuss the interplay between superconductivity
and the nodal and antinodal SC quasiparticle coherences in cuprate
superconductors under this kinetic energy driven SC mechanism, and
show that the s-wave SC transition temperature is heavily
suppressed by the antinodal SC quasiparticle coherence, while the
d-wave SC transition temperature is enhanced, therefore the
antinodal SC quasiparticle coherence plays a more crucial role for
superconductivity in doped cuprates.

Very soon after the discovery of superconductivity in doped
cuprates, Anderson \cite{anderson} suggested that the essential
physics of doped cuprates is contained in the $t$-$J$ model on a
square lattice,
\begin{eqnarray}
H&=&-t\sum_{i\hat{\eta}\sigma}C^{\dagger}_{i\sigma}
C_{i+\hat{\eta}\sigma}+t'\sum_{i\hat{\tau}\sigma}
C^{\dagger}_{i\sigma}C_{i+\hat{\tau}\sigma} \nonumber\\
&+&\mu\sum_{i\sigma} C^{\dagger}_{i\sigma}C_{i\sigma}
+J\sum_{i\hat{\eta}}{\bf S}_{i} \cdot {\bf S}_{i+\hat{\eta}},
\end{eqnarray}
where $\hat{\eta}=\pm\hat{x},\pm \hat{y}$, $\hat{\tau}=\pm\hat{x}
\pm\hat{y}$, $C^{\dagger}_{i\sigma}$ ($C_{i\sigma}$) is the
electron creation (annihilation) operator, ${\bf S}_{i}=
C^{\dagger}_{i}{\vec\sigma} C_{i}/2$ is spin operator with
${\vec\sigma}=(\sigma_{x},\sigma_{y},\sigma_{z})$ as Pauli
matrices, and $\mu$ is the chemical potential. This $t$-$J$ model
is subject to an important local constraint $\sum_{\sigma}
C^{\dagger}_{i\sigma}C_{i\sigma}\leq 1$ to avoid the double
occupancy, which can be treated properly in analytical
calculations within the CSS fermion-spin theory \cite{feng2},
where the constrained electron operators are decoupled as
$C_{i\uparrow}= h^{\dagger}_{i\uparrow} S^{-}_{i}$ and
$C_{i\downarrow}= h^{\dagger}_{i\downarrow} S^{+}_{i}$, with the
gauge invariant {\it spinful} fermion operator $h_{i\sigma}=
e^{-i\Phi_{i\sigma}} h_{i}$ describes the charge degree of freedom
together with some effects of the spin configuration
rearrangements due to the presence of the doped hole itself
(dressed holon), while the spin operator $S_{i}$ describes the
spin degree of freedom (spin), then the electron local constraint
for the single occupancy is satisfied in analytical calculations.
In this CSS fermion-spin representation, the low-energy behavior
of the $t$-$J$ model can be expressed as \cite{feng5},
\begin{eqnarray}
H&=&t\sum_{i\hat{\eta}}(h^{\dagger}_{i+\hat{\eta}\uparrow}
h_{i\uparrow}S^{+}_{i}S^{-}_{i+\hat{\eta}}+
h^{\dagger}_{i+\hat{\eta}\downarrow}h_{i\downarrow}S^{-}_{i}
S^{+}_{i+\hat{\eta}})\nonumber \\
&-&t'\sum_{i\hat{\tau}}(h^{\dagger}_{i+\hat{\tau}\uparrow}
h_{i\uparrow}S^{+}_{i}S^{-}_{i+\hat{\tau}}+
h^{\dagger}_{i+\hat{\tau}\downarrow}h_{i\downarrow}S^{-}_{i}
S^{+}_{i+\hat{\tau}}) \nonumber \\
&-&\mu\sum_{i\sigma}h^{\dagger}_{i\sigma} h_{i\sigma}+J_{{\rm
eff}}\sum_{i\hat{\eta}}{\bf S}_{i}\cdot {\bf S}_{i+\hat{\eta}},
\end{eqnarray}
with $J_{{\rm eff}}=(1-\delta)^{2}J$, and $\delta=\langle
h^{\dagger}_{i\sigma}h_{i\sigma}\rangle=\langle h^{\dagger}_{i}
h_{i}\rangle$ is the doping concentration.

Since ARPES measurements \cite{shen1} have shown that in the real
space the gap function and pairing force have a range of one
lattice spacing, then the order parameter for the electron Cooper
pair in the doped regime without AFLRO can be expressed as
\cite{feng1,feng3}, $\Delta=\langle C^{\dagger}_{i\uparrow}
C^{\dagger}_{i+\hat{\eta}\downarrow} -C^{\dagger}_{i\downarrow}
C^{\dagger}_{i+\hat{\eta}\uparrow}\rangle=-\chi_{1}\Delta_{h}$,
with the spin correlation function $\chi_{1}=\langle S^{+}_{i}
S^{-}_{i+\hat{\eta}}\rangle$ and dressed holon pairing order
parameter $\Delta_{h}= \langle h_{i+\hat{\eta}\downarrow}
h_{i\uparrow}-h_{i+\hat{\eta}\uparrow} h_{i\downarrow}\rangle$,
which shows that the SC order parameter for the electron Cooper
pair is related to the dressed holon pairing amplitude, and is
proportional to the number of doped holes, and not to the number
of electrons. In this case, we \cite{feng1,feng3} have shown
within the Eliashberg's strong coupling theory \cite{eliashberg}
that the dressed holon-spin interaction can induce the dressed
holon pairing state (then the electron Cooper pairing state) by
exchanging the spin excitations in the higher power of the doping
concentration, where the full dressed holon BCS type diagonal and
off-diagonal Green's functions of the $t$-$J$ model (2) have been
obtained as \cite{feng5},
\begin{eqnarray}
g(k)&=&Z_{hF}({\bf k})\left ({U^{2}_{h{\bf k}}\over
i\omega_{n}-E_{h{\bf k}}}+{V^{2}_{h{\bf k}}\over i\omega_{n}
+E_{h{\bf k}}}\right ),\\
\Im^{\dagger}(k)&=&-{\bar{\Delta}_{hZ}({\bf k}) \over 2E_{h{\bf
k}}}\left ({Z_{hF}({\bf k})\over i\omega_{n}-E_{h{\bf k}}}-
{Z_{hF}({\bf k})\over i\omega_{n}+ E_{h{\bf k}}} \right ),
\end{eqnarray}
respectively, where the four-vector notation $k=({\bf k},
i\omega_{n})$, $U^{2}_{h{\bf k}}=(1+\bar{\xi_{{\bf k}}}/E_{h{\bf
k}})/2$, $V^{2}_{h{\bf k}}=(1-\bar{\xi_{{\bf k}}}/E_{h{\bf
k}})/2$, $\bar{\Delta}_{hZ}({\bf k})=Z_{hF}({\bf
k})\bar{\Delta}_{h}({\bf k})$, $E_{h{\bf k}}= \sqrt
{\bar{\xi^{2}_{{\bf k}}}+ \mid\bar{\Delta}_{hZ}({\bf k})
\mid^{2}}$, $\bar{\xi_{{\bf k}}} =Z_{hF}({\bf k})[\xi_{{\bf k}}
+\Sigma^{(h)}_{1e}({\bf k})]$, while the static limit dressed
holon effective gap function $\bar{\Delta}_{h}({\bf k})
=\Sigma^{(h)}_{2}({\bf k},\omega) \mid_{\omega=0}$ and
quasiparticle coherent weight $Z^{-1}_{hF}({\bf k})=
1-\Sigma^{(h)}_{1o}({\bf k},\omega) \mid_{\omega=0}$, with the
dressed holon self-energy functions from the spin bubble
\cite{feng3,feng5},
\begin{eqnarray}
\Sigma^{(h)}_{1}(k)&=&{1\over N^{2}}\sum_{{\bf p,p'}}
(Zt\gamma_{{\bf p+p'+k}}- Zt'\gamma_{{\bf p+p'+k}}')^{2}
\nonumber \\
&\times&{1\over\beta}\sum_{ip_{m}}g(p+k)\Pi({\bf p'},p), \\
\Sigma^{(h)}_{2}(k)&=&{1\over N^{2}}\sum_{{\bf p,p'}}(Zt
\gamma_{{\bf p+p'+k}}-Zt'\gamma_{{\bf p+p'+k}}')^{2}
\nonumber \\
&\times&{1\over \beta}\sum_{ip_{m}}\Im (-p-k)\Pi({\bf p'},p),
\end{eqnarray}
where $\Pi({\bf p'},p)=(1/\beta)\sum_{ip'_{m}}D^{(0)}(p')D^{(0)}
(p'+p)$, $\gamma_{{\bf k}}=(1/Z)\sum_{\hat{\eta}}e^{i{\bf k}\cdot
\hat{\eta}}$, $\gamma_{{\bf k}}'=(1/Z)\sum_{\hat{\tau}}e^{i{\bf k}
\cdot\hat{\tau}}$, $Z$ is the number of the nearest neighbor or
second-nearest neighbor sites, $p=({\bf p},ip_{m})$, $p'=({\bf
p'}, ip'_{m})$, $\Sigma^{(h)}_{1o}(k)$ and $\Sigma^{(h)}_{1e}(k)$
are the symmetric and antisymmetric parts of
$\Sigma^{(h)}_{1}(k)$, while the mean-field (MF) spin Green's
function, $D^{(0)-1}(p)= [(ip_{m})^{2}-\omega_{{\bf p}}^{2}]
/B_{{\bf p}}$, with $B_{{\bf p}}$, the MF spin excitation spectrum
$\omega_{{\bf p}}$, and MF dressed holon excitation spectrum
$\xi_{{\bf k}}$ have been given in Ref. \cite{feng5}.

Although $Z_{hF}({\bf k})$ still is a function of ${\bf k}$, the
wave vector dependence may be unimportant, since ARPES experiments
have shown that around the nodal and antinodal points contain the
essentials of the whole low energy quasiparticle spectrum
\cite{shen,fink}. Therefore in the following discussions, we only
study the effects of the nodal and antinodal SC quasiparticle
coherences on superconductivity, i.e., $Z^{(N)}_{hF}=Z_{hF}({\bf
k})\mid_{{\bf k}=[\pi/2,\pi/2]}$ at the nodal point, and
$Z^{(A)}_{hF}=Z_{hF}({\bf k})\mid_{{\bf k}=[\pi,0]}$ at the
antinodal point. On the other hand, for the understanding of the
different influences of the SC quasiparticle coherence on the
s-wave and d-wave SC-states, we consider,
$\bar{\Delta}^{(s)}_{hZ}({\bf k})= \bar{\Delta}^{(s)}_{hZ}
\gamma^{(s)}_{{\bf k}}$, with $\gamma^{(s)}_{{\bf k}}=\gamma_{{\bf
k}}=({\rm cos}k_{x}+{\rm cos} k_{y})/2$ for the s-wave pairing,
and $\bar{\Delta}^{(d)}_{hZ} ({\bf k})= \bar{\Delta}^{(d)}_{hZ}
\gamma^{(d)}_{{\bf k}}$, with $\gamma^{(d)}_{{\bf k}}=({\rm cos}
k_{x}-{\rm cos}k_{y})/2$ for the d-wave pairing. In this case, the
dressed holon effective gap parameter and quasiparticle coherent
weight satisfy the following two equations \cite{feng3,feng5},
\begin{eqnarray}
1&=&{1\over\bar{\Delta}^{(a)}_{h}}{4\over N}\sum_{{\bf k}}
\gamma^{(a)}_{{\bf k}}\Sigma^{(h)}_{2}({\bf k},\omega)
\mid_{\omega=0}, \\
Z^{(\alpha)-1}_{hF}&=&1-\Sigma^{(h)}_{1o}({\bf k}_{\alpha},
\omega)\mid_{\omega=0},
\end{eqnarray}
respectively, where $a=s, d$, $\alpha=N, A$, and ${\bf k}_{\alpha}
={\bf k}_{N}, {\bf k}_{A}$, with ${\bf k}_{N}=[\pi/2,\pi/2]$ and
${\bf k}_{A}=[\pi,0]$. These two equations must be solved
simultaneously with other self-consistent equations
\cite{feng3,feng5}, then all order parameters and chemical
potential $\mu$ are determined by the self-consistent calculation.

With the help of the above discussions, we now can obtain the
dressed holon pair gap parameter in terms of the holon
off-diagonal Green's function (4) as $\Delta^{(a)}_{h}=(2/N)
\sum_{{\bf k}}[\gamma^{(a)}_{{\bf k}}]^{2}{Z^{(\alpha)}_{hF}
\bar{\Delta}^{(a)}_{hZ}{\rm tanh}[{1\over 2}\beta E_{h{\bf k}}]
/E_{h{\bf k}}}$. This dressed holon pairing state originating from
the kinetic energy terms by exchanging the spin excitations also
leads to form the electron Cooper pairing state
\cite{feng1,feng3}, where the electron quasiparticle coherent
weight and SC gap function are obtained from the electron diagonal
and off-diagonal Green's functions $G(i-j,t-t')=\langle\langle
C_{i\sigma}(t); C^{\dagger}_{j\sigma}(t')\rangle\rangle$ and
$\Gamma^{\dagger} (i-j,t-t')=\langle\langle
C^{\dagger}_{i\uparrow}(t); C^{\dagger}_{j\downarrow}(t')
\rangle\rangle$, which are the convolutions of the spin Green's
function and dressed holon diagonal and off-diagonal Green's
functions, respectively, and reflect CSR \cite{anderson1}. This
CSR transfers the dressed holon BCS type diagonal and off-diagonal
Green's functions (3) and (4) into the corresponding electron BCS
type diagonal and off-diagonal Green's functions, then the nature
of the SC quasiparticle coherence is described by the simple BCS
formalism \cite{feng5}, although the pairing mechanism is driven
by the kinetic energy by exchanging the spin excitations.
Following our previous discussions \cite{feng3,feng5}, we can
obtain $G(k)$ and $\Gamma^{\dagger}(k)$, then the electron
quasiparticle coherent weight and effective SC gap parameter are
evaluated as $Z^{(\alpha)}_{F}\approx Z^{(\alpha)}_{hF} /2$ and
$\bar{\Delta}^{(a)}\approx-\chi_{1}\bar {\Delta}^{(a)}_{h}$,
respectively.

\begin{figure}
\begin{center}
\includegraphics*[height=4cm,width=8cm]{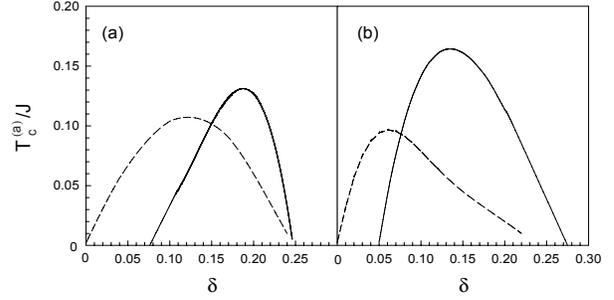}
\end{center}
\caption{The SC transition temperature $T^{(a)}_{c}$ as a function
of doping in the d-wave (solid line) and s-wave (dashed line)
cases for $t/J=2.5$ and $t'/J=0.3$ with the SC quasiparticle
coherence at (a) the nodal and (b) antinodal points.}
\end{figure}

As we \cite{feng1,feng3} have shown that the SC transition
temperature $T^{(a)}_{c}$ occurring in the case of the SC gap
parameter $\Delta^{(a)}=0$ is identical to the dressed holon pair
transition temperature occurring in the case of the dressed holon
pairing gap parameter $\Delta^{(a)}_{h}=0$. In Fig. 1, we plot the
SC transition temperature $T^{(a)}_{c}$ as a function of doping in
the d-wave (solid line) and s-wave (dashed line) cases for
$t/J=2.5$ and $t'/J=0.3$ with the SC quasiparticle coherence at
(a) the nodal and (b) antinodal points. It is shown that for the
s-wave symmetry, the maximal SC transition temperature
T$^{(s)}_{c}$ occurs around a particular doping concentration, and
then decreases in both lower doped and higher doped regimes, while
for the d-wave symmetry, the maximal SC transition temperature
T$^{(d)}_{c}$ occurs around the optimal doping, and then decreases
in both underdoped and overdoped regimes. However, the s-wave SC
transition temperature is heavily suppressed by the antinodal
quasiparticle coherence, while the d-wave SC transition
temperature is enhanced. Since the experimental results
\cite{shen,matsui,ding} have shown that the SC-state in doped
cuprates has the d-wave symmetry in a wide range of doping, then
in this sense, the antinodal quasiparticle coherence plays a more
crucial role for superconductivity. These results also are
consistent with the experimental evidence that the most
contributions of the electronic states for cuprate superconductors
come from the antinodal point \cite{shen,fink}.

\begin{figure}
\begin{center}
\includegraphics*[height=4cm,width=8cm]{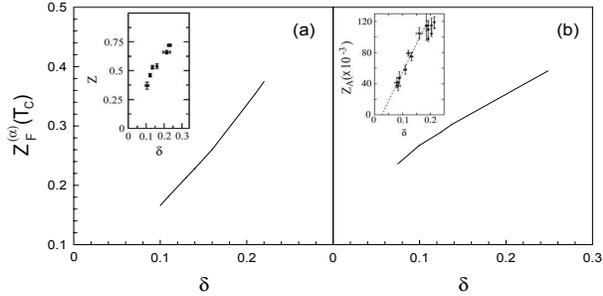}
\end{center}
\caption{(a) The nodal and (b) antinodal quasiparticle coherent
weights $Z^{(\alpha)}_{F}(T_{c})$ as a function of doping for
$t/J=2.5$ and $t'/t=0.3$. Inset: the corresponding experimental
results of the nodal \protect\cite{johnson} and antinodal
\protect\cite{ding} quasiparticle coherent weights.}
\end{figure}

The essential physics of superconductivity in the present case is
the same as that in our previous discussions \cite{feng3,feng5}.
The antisymmetric part of the self-energy $\Sigma^{(h)}_{1o}({\bf
k})$ (then $Z^{(\alpha)}_{F}$) describes the dressed holon (then
electron) quasiparticle coherence, and therefore
$Z^{(\alpha)}_{F}$ is closely related to the SC quasiparticle
density, while the self-energy $\Sigma^{(h)}_{2}({\bf k})$
describes the effective dressed holon (then electron) pairing gap
function. In particular, $Z^{(\alpha)}_{F}$ is doping dependent.
To show this point clearly, we plot (a) the nodal and (b)
antinodal SC quasiparticle coherent weights
$Z^{(\alpha)}_{F}(T_{c})$ as a function of doping for $t/J=2.5$
and $t'/t=0.3$ in Fig. 2 in comparison with the corresponding
experimental results of the nodal \cite{johnson} and antinodal
\cite{ding} SC quasiparticle coherent weights (inset). As seen
from Fig. 2, $Z^{(\alpha)}_{F}$ grows linearly with doping, i.e.,
$Z^{(\alpha)}_{F}\propto \delta$, which together with the SC gap
parameter \cite{feng3,feng5} show that only $\delta$ number of the
coherent doped carriers are recovered in the SC-state, consistent
with the picture of a doped Mott insulator with $\delta$ holes
\cite{anderson}. Since the SC-order is established through an
emerging SC quasiparticle \cite{ding}, therefore the SC-order is
controlled by both SC gap function and quasiparticle coherence,
and is reflected explicitly in the self-consistent equations (7)
and (8), this leads to that the SC transition temperature
increases with increasing doping in the underdoped regime, and
reaches a maximum in the optimal doping, then decreases in the
overdoped regime.

In summary, we have discussed the interplay between
superconductivity and the nodal and antinodal SC quasiparticle
coherences in cuprate superconductors under the kinetic energy
driven SC mechanism. It is shown that the s-wave SC transition
temperature is heavily suppressed by the antinodal SC
quasiparticle coherence, while the d-wave SC transition
temperature is enhanced, therefore the antinodal SC quasiparticle
coherence plays a more crucial role for superconductivity.


This work was supported by the National Natural Science Foundation
of China under Grant Nos. 10125415 and 90403005, and the Grant
from the Ministry of Science and Technology of China under Grant
No. 2006CB601002.

\end{document}